\magnification=\magstep1
\parindent=0pt
\parskip=6pt
\openup 1\jot
\font\titlefont = cmbx10 scaled \magstep3
\font\authorfont = cmbx10 scaled \magstep1
\input amssym.def
\input amssym.tex
\def\a{\alpha}
\def\b{\beta}
\def\d{\delta}
\def\da{\delta_{\alpha}}
\def\db{\delta_{\beta}}
\def\td{{\tilde \delta}}
\def\tda{{\tilde \delta}_{\alpha}}
\def\tdb{{\tilde \delta}_{\beta}}
\def\D{\Delta}

\def\e{\epsilon}
\def\p{\psi}
\def\l{\lambda}
\def\m{\mu}
\def\n{\nu}
\def\s{\sigma}
\def\t{\tau}
\def\adot{\dot a}
\def\bdot{\dot b}
\def\pdot{\dot \psi}
\def\xdot{\dot x}
\def\Qa{Q_{\alpha}}
\def\Qb{Q_{\beta}}
\def\tQ{\tilde Q}
\def\tQa{{\tilde Q}_{\alpha}}
\def\tQb{{\tilde Q}_{\beta}}
\def\half{{1 \over 2}}
\rightline {DAMTP 1998 -- 39}
\vskip 20pt
\centerline {\titlefont Deconstructing Supersymmetry}
\vskip 20pt
\centerline{{\authorfont N.S. Manton}\footnote {${}^*$} {email:
N.S.Manton@damtp.cam.ac.uk}}
\vskip 15pt
\centerline{\it Department of Applied Mathematics and Theoretical
Physics} 
\centerline{\it University of Cambridge} 
\centerline{\it Silver Street, Cambridge CB3 9EW, England}
\vskip 30pt

{\bf Abstract}
\vskip 5pt
Two supersymmetric classical mechanical systems are
discussed. Concrete realizations are obtained by supposing that the
dynamical variables take values in a Grassmann algebra with two
generators. The equations of motion are explicitly solved. 
A genuine Lie group, the supergroup, generated by supersymmetries and
time translations, is found to act on the space of solutions. For each
system, the solutions with zero energy need to be constructed
separately. For these Bogomolny-type solutions, the orbit of the
supergroup is smaller than in the generic case.


\vfill\eject
{\bf I. Introduction}

Supersymmetry is one of the most powerful ideas in theoretical
physics, combining bosonic and fermionic fields into a unified
framework. Most supersymmetric theories are defined by a Lagrangian,
from which the classical field equations are derived. However the
meaning of the fermionic fields in such equations is not always clear,
because they need to be anticommuting. Moreover, there are 
usually sources for the bosonic fields which are bilinear in the 
fermionic fields, and such sources are not ordinary functions. So an 
interpretation of the bosonic fields as ordinary functions fails.

In fact, the formalism for making sense of classical supersymmetric
theories is readily available, but perhaps not sufficiently
appreciated by theoretical physicists. It is the substance of the book
by de Witt [1], and is also repeatedly mentioned in the earlier chapters
of Freund's book [2]. Fields in a supersymmetric field theory must take
their values in a Grassmann algebra $B$. $B$ is the direct sum of an
even part $B_e$ and an odd part $B_o$. The bosonic fields are valued
in $B_e$, and the fermionic fields in $B_o$. It is necessary to
decide which algebra $B$ to work with. $B$ can have a finite number,
$n$, of generators, or an infinite number, and the content of the
theory will depend on the choice. With $n$ generators, a scalar
bosonic field is represented by $2^{n-1}$ ordinary functions, 
and by an infinite number if $B$ is infinitely generated. This
is rather daunting. However, we shall choose $n=2$ in what follows,
and the resulting equations are quite manageable. (The choice $n=1$
leads to trivial equations.)

Mechanical models, with bosonic and fermionic dynamical variables
taking values in a Grassmann algebra, and depending only on time, 
were investigated by Casalbuoni [3] and by Berezin and
Marinov [4], although not solved except in very simple
cases. Supersymmetry constrains the structure of such models. 
We analyse two supersymmetric mechanical models below. We
present the Lagrangian and equations of motion, their symmetries
and the associated conserved quantities, and proceed to find the explicit
form of the general solution of the equations of motion. 
We believe that this has not been
done before. The possibility of constructing general solutions of the
nonlinear coupled ODE's shows the power of the supersymmetry of
these models. From the supersymmetry algebra we construct a genuine
Lie algebra of infinitesimal symmetries, which generates a genuine Lie
group of symmetries of the dynamics. This group, which depends on $n$,
we call the supergroup. The solutions depend on a number of 
constants of integration, and we comment on the extent to which 
the supergroup relates solutions with different values
of these constants.

For each of these models, the solutions with zero energy need to
be constructed independently. Here, one of the bosonic equations of
motion reduces to a first-order Bogomolny-type equation [5]. The
solution space is
still acted on by the supergroup, but the orbit is of
lower dimension than in the generic case. This feature of
Bogomolny equations is not unfamiliar, but the complete solution of
the equations of motion, including the fermionic variables, is perhaps novel.

Section II discusses the $N=2$ supersymmetric mechanics of a particle
moving in one dimension, subject to a potential. The model is a variant
of the one whose quantized version was analysed by
Witten [6]. Section III is concerned
with the zero energy, Bogomolny case. Section IV
discusses the $N=1$ supersymmetric mechanics of a particle
moving in one dimension. Again the model is a variant of the standard
one, as the Lagrangian depends on a constant odd parameter. We conclude 
in Section V with some comments
on the analysis, and on potential generalizations of this work.

{\bf II. $N=2$ Supersymmetric Mechanics}

Consider the following $N=2$ supersymmetric Lagrangian [6], [1, \S 5.7]
$$
L = \half {\xdot}^2 + \half U(x)^2 + \half\pdot_1 \p_1 
- \half\pdot_2\p_2 + U'(x)\p_1\p_2 \ .
\eqno(2.1)
$$
This describes the supersymmetric mechanics of a particle moving in one
dimension in a potential $-U^2$.
$x(t)$ is bosonic (i.e. commuting) and $\p_1(t)$ and $\p_2(t)$ are
fermionic (i.e. anticommuting) variables. Thus $x$ is valued in
$B_e$, whereas $\p_1$ and $\p_2$ are valued in $B_o$. Any function of $x$,
e.g. $U(x)$, commutes with $x$. Such functions are defined as
polynomials or power series with real coefficients. If $U(x) = x^p$,
with $p$ a positive integer, then $U'(x) = px^{p-1}$, with the obvious
extension to polynomials and power series. An overdot
denotes the derivative with respect to time $t$. $\xdot$ commutes
with $x$, and similarly, $\pdot_1$ and $\pdot_2$ anticommute
with both $\p_1$ and $\p_2$; hence the dynamics is classical, rather
than quantized. Note that the
terms $\pdot_1 \p_1$ and $\pdot_2 \p_2$ are not total
time derivatives.

The Lagrangian $L$ may be obtained by dimensional reduction of the
$1+1$ dimensional $N=1$ supersymmetric field theory with Lagrangian
density
$$
{\cal L} = \half \partial_+\Phi \partial_-\Phi - \half U(\Phi)^2  
+ {i \over 2}\p_1\partial_-\p_1 + {i \over 2}\p_2\partial_+\p_2 
+ i{{dU} \over {d\Phi}}\p_1\p_2 \ ,
\eqno(2.2)
$$
where $\partial_+$ and $\partial_-$ are the standard light cone derivatives.
By assuming that all fields are independent of the spatial coordinate,
then absorbing certain factors of ${\sqrt i}$ etc. in the fields and
potential, and
finally writing $\Phi$ as $x$, we recover the expression (2.1). The
density (2.2) is real in a certain sense related to quantization, but
for our purposes the manifestly real expression (2.1) is a more
convenient Lagrangian to discuss.

To obtain the equations of motion we calculate the formal variation
$\D L$ due to variations $\D x$, $\D\p_1$ and $\D\p_2$. We combine
$\D\xdot$, $\D\pdot_1$ and $\D\pdot_2$ into total time derivative
terms, which are ignored,
then move $\D x$, $\D\p_1$ and $\D\p_2$ to the left
in each term. The result is
$$
\D L = \D x(- {\ddot x} + UU' + U''\p_1\p_2) + \D \p_1(-\pdot_1 + U'\p_2) +
\D \p_2(\pdot_2 - U'\p_1) \ ,
\eqno(2.3)
$$
so the equations of motion are
$$
\eqalignno{
{\ddot x} &= UU' + U''\p_1\p_2 &(2.4a) \cr
\pdot_1 &= U'\p_2 &(2.4b) \cr
\pdot_2 &= U'\p_1 \ . &(2.4c) \cr
}
$$

The Lagrangian has two supersymmetries. The first is defined by the
variations
$$
\d x = \e \p_1 \ , \ \d \p_1 = \e \xdot \ , \ \d \p_2 = \e U \ ,
\eqno(2.5)
$$
where $\e$ is an arbitrary infinitesimal constant in $B_o$. It is
easily shown that the variation of $L$ is a total time-derivative
$$
\d L = \e {d \over dt}(\half \xdot \p_1 + \half U \p_2 )
\eqno(2.6)
$$
using ${\dot U} = U' \xdot$. The usual Noether method gives the
conserved quantity
$$
Q = \xdot \p_1 - U \p_2 \ .
\eqno(2.7)
$$
The conservation of $Q$ is easily verified using the equations of
motion:
$$
\eqalign{
{\dot Q} &= {\ddot x}\p_1 + \xdot \pdot_1 - U' \xdot \p_2 - U\pdot_2 \cr
&= UU'\p_1 + U''\p_1\p_2\p_1 + \xdot U' \p_2  - U' \xdot \p_2 -
UU'\p_1 \cr 
&= 0}
\eqno(2.8)
$$
since $\p_1\p_2\p_1 = -\p_1\p_1\p_2 = 0$.
The second supersymmetry is defined by the variations
$$
\td x = \e \p_2 \ , \ \td \p_2 = -\e \xdot \ , \ \td \p_1 = -\e U \ ,
\eqno(2.9)
$$
and leads to the conserved quantity
$$
\tQ = \xdot \p_2 - U \p_1 \ .
\eqno(2.10)
$$

The supersymmetries relate different solutions of the equations of
motion. To see this, consider the linearized variations of the
equations (2.4)
$$
\eqalignno{
{\ddot {(\D x)}} &= (UU')' \ \D x + U'''\D x \ \p_1 \p_2 + U''\D \p_1\p_2 +
U''\p_1\D \p_2 &(2.11a) \cr
{\dot {(\D \p_1)}} &= U''\D x \ \p_2 + U'\D \p_2 &(2.11b) \cr
{\dot {(\D \p_2)}} &= U''\D x \ \p_1 + U'\D \p_1 &(2.11c) \cr
}
$$
and assume that $x$, $\p_1$ and $\p_2$ satisfy (2.4).
The linear equations (2.11) are satisfied by setting 
$\D = \d$ or $\D = \td$, and using the
variations defined in (2.5) and (2.9). Later, we shall see more
concretely, and not just in the linearized approximation, how
supersymmetry relates different solutions.

Since the Lagrangian (2.1) does not depend explicitly on time, we
expect a conserved energy, associated with time translation
symmetry. The coefficient of the time translation is an arbitrary
infinitesimal element of $B_e$. The energy is
$$
H = \half \xdot^2 - \half U^2 - U'\p_1\p_2 \ ,
\eqno(2.12)
$$
and its conservation is easily checked using the equations of motion.

We now simplify matters, and make the model more concrete, by
supposing that the Grassmann algebra $B$ is
generated by just two elements $\a , \b$ satisfying
$$
\a^2 = 0 \ , \ \b^2 = 0 \ , \ \a\b + \b\a = 0 \ .
\eqno(2.13)
$$
A basis for the algebra is $\{ 1,\a,\b,\a\b \}$, and it follows from (2.13)
that $(\a\b)^2 = 0$. There is a matrix realization of these
relations, although we will not use it. Let $\{ \gamma^{\mu}: 1 \leq
\mu \leq 4 \}$ denote Dirac matrices in four Euclidean
dimensions, and set $\a = \gamma^1 + i\gamma^2 , \b = \gamma^3 +
i\gamma^4$.

Let us write the dynamical variables in component form as
$$ 
\eqalignno{
x(t) &= x_0(t) + x_1(t)\a\b &(2.14a) \cr
\p_1(t) &= a_1(t)\a + b_1(t)\b &(2.14b) \cr
\p_2(t) &= a_2(t)\a + b_2(t)\b &(2.14c) \cr
}
$$
where $x_0,x_1,a_1,b_1,a_2,b_2$ are ordinary functions of time. The
``body'', $x_0(t)$, can be regarded as classical.

Any positive power of $x$ has the expansion
$$
x^n = x_0^n + nx_0^{n-1}x_1\a\b
\eqno(2.15)
$$
which extends to an arbitrary function of $x$ as
$$
U(x) = U(x_0) + U'(x_0)x_1\a\b
\eqno(2.16)
$$
where $U'(x_0)$ denotes the usual derivative of $U(x_0)$ with respect
to $x_0$. Henceforth, if the argument of $U$ and its derivatives
is not shown, it is $x_0$, with $x_0$ itself a function of $t$. 
The Lagrangian is the even function $L = L_0 + L_1\a\b$,
where
$$
\eqalignno{
L_0 &= \half {\xdot}_0^2 + \half U^2 &(2.17a) \cr
L_1 &= \xdot_0 \xdot_1 + UU' x_1 + \adot_1 b_1 - \adot_2 b_2
+ U'(a_1 b_2 - a_2 b_1) \ . &(2.17b) \cr
}
$$

Substituting (2.14) into (2.4), we obtain the equations of motion 
for the components
$$
\eqalignno{
{\ddot x}_0 &= UU' &(2.18a) \cr
{\ddot x}_1 &= (UU')' x_1 + U'' (a_1 b_2 - a_2 b_1) &(2.18b) \cr
\adot_1 &= U' a_2 &(2.18c) \cr
\adot_2 &= U' a_1 &(2.18d) \cr
\bdot_1 &= U' b_2 &(2.18e) \cr
\bdot_2 &= U' b_1 \ . &(2.18f) \cr
}
$$
These equations can also be derived as the variational equations of
$L_0$ and $L_1$. In fact, surprisingly, they can all be derived from
$L_1$ alone, as the equation of motion for $x_0$, obtained from $L_0$,
is the same as the equation obtained from $L_1$ by varying $x_1$.

There are a host of symmetries and conservation laws associated with
the component form of the system. Some of these relate to
supersymmetry. We may define two supersymmetry variations $\da$ and
$\db$, associated with $\d$. $\da$ is defined, following (2.5), by
$$
\da x = \e \a \p_1 \ , \ \da \p_1 = \e \a \xdot \ , \ \da \p_2 = \e \a
U(x) \ ,
\eqno(2.19)
$$
where $\e$ is now infinitesimal and real, and $\db$ similarly by
replacing $\a$ by $\b$. In components, the first of these variations
becomes
$$
\da (x_0 + x_1\a\b) = \e b_1 \a\b
\eqno(2.20)
$$
so $\da x_0 = 0$ and $\da x_1 = \e b_1$. Similarly, by expanding out,
we find the complete set of variations
$$
\eqalignno{
\da x_1 &= \e b_1 \ , \ \da a_1 = \e \xdot_0 \ , \ \da a_2 = \e U 
&(2.21a) \cr
\db x_1 &= -\e a_1 \ , \ \db b_1 = \e \xdot_0 \ , \ \db b_2 = \e U \ ,
&(2.21b) \cr
}
$$
with all other variations, e.g. $\db a_1$, vanishing. The
supersymmetry $\td$ leads similarly to the two independent sets of variations
$$
\eqalignno{
\tda x_1 &= \e b_2 \ , \ \tda a_1 = -\e U \ , \ 
\tda a_2 = -\e \xdot_0 &(2.22a) \cr
\tdb x_1 &= -\e a_2 \ , \ \tdb b_1 = -\e U \ , \ 
\tdb b_2 = -\e \xdot_0 \ . &(2.22b) \cr
}
$$
$x_0$, and hence $L_0$ is unchanged by all these variations. 

It is easy to verify that all four sets of variations $\da ,\db ,\tda
,\tdb$ are Noether symmetries of the Lagrangian $L_1$, giving total
time derivatives. For example
$$
\eqalign{
\da L_1 &= \e (\xdot_0 \bdot_1 + UU' b_1 + {\ddot x}_0 b_1 - U' \xdot_0
b_2 + U' (\xdot_0 b_2 - Ub_1)) \cr 
&= \e {d \over dt} (\xdot_0 b_1) \ .}
\eqno(2.23) 
$$
In the usual way, we obtain the conserved Noether
charges
$$
\eqalignno{
\Qa &= \xdot_0 b_1 - Ub_2 &(2.24a) \cr
\Qb &= -\xdot_0 a_1 + Ua_2 &(2.24b) \cr
\tQa &= \xdot_0 b_2 - Ub_1 &(2.24c) \cr
\tQb &= -\xdot_0 a_2 + Ua_1 \ , &(2.24d) \cr
}
$$
and may verify their conservation using the equations
of motion (2.18). Of course, these charges are just the components of
the supersymmetry charges we found earlier, although with labels
switched, namely
$$
\eqalignno{
Q &= -\Qb \a + \Qa \b &(2.25a) \cr
\tQ &= -\tQb \a + \tQa \b \ . &(2.25b) \cr
}
$$
Both $L_0$ and $L_1$ are invariant under time translations, leading to
the conservation of two energies
$$
\eqalignno{
H_0 &= \half {\xdot}_0^2 - \half U^2 &(2.26a) \cr
H_1 &= \xdot_0 \xdot_1 - UU' x_1
- U' (a_1 b_2 - a_2 b_1) \ . &(2.26b) \cr
}
$$
The conserved energy we found earlier is $H = H_0 + H_1 \a\b$.

There is a further symmetry, a mini-time-translation symmetry, arising
from an infinitesimal time translation with coefficient proportional
to $\a\b$
$$
\D x = \e \a\b \xdot \ , \ \D \p_1 = \e \a\b \pdot_1 \ , \ \D \p_2 = \e \a\b
\pdot_2 \ .
\eqno(2.27)
$$
Expanding out in components, we find a single nonzero variation
$$
\D x_1 = \e \xdot_0 \ .
\eqno(2.28)
$$
The associated variation of $L_1$ is
$$
\eqalign{
\D L_1 &= \e (\xdot_0 {\ddot x}_0 + UU' \xdot_0 ) \cr 
&= \e {d \over dt} (\half \xdot_0^2 + \half U^2) \ ,}
\eqno(2.29) 
$$
and the conserved quantity is
$$
\half\xdot_0^2 - \half U^2 \ ,
\eqno(2.30)
$$
which is $H_0$. So we see that the equations of motion and both
conserved energies, and all four components of the supersymmetry
charges, can be derived from $L_1$.

There are yet more symmetries which mix the functions $a_1, a_2, b_1,
b_2$. The combined variations 
$$
\D a_1 = \e b_1 \ , \ \D a_2 = \e b_2 
\eqno(2.31)
$$
leave $L_1$ invariant, as do the variations
$$
\D b_1 = \e a_1 \ , \ \D b_2 = \e a_2 \ .
\eqno(2.32)
$$
Finally, $L_1$ is invariant under
$$
\D a_1 = \e a_2 \ , \ \D a_2 = \e a_1 \ , \D b_1 = \e b_2 \ , \ 
\D b_2 = \e b_1 \ .
\eqno(2.33)
$$
These symmetries imply that
$$
\eqalignno{
R_a &= \half (b_1^2 - b_2^2) &(2.34a) \cr
R_b &= \half (a_1^2 - a_2^2) &(2.34b) \cr
R &= a_1 b_2 - a_2 b_1 &(2.34c) \cr
}
$$
are all conserved.

The conservation of $R$ can also be understood from the symmetry of
the original Lagrangian $L$ under the infinitesimal variations
$$
\D \p_1 = \e \p_2 \ , \ \D \p_2 = \e \p_1
\eqno(2.35)
$$
with $\e$ real, which implies the conservation of $\p_1 \p_2$.

We turn now to the solution of the coupled equations (2.18). We start with
the equation for $x_0$. This is the classical equation of the model
without fermionic variables. It has the first integral
$$
\xdot_0^2 - U^2 = 2E \ ,
\eqno(2.36)
$$
where $H_0 = E$ is the conserved energy, hence
$$
\xdot_0 = (2E + U^2)^{\half} \ .
\eqno(2.37)
$$
The solution in integral form is
$$
\int_{X_0}^{x_0} {{dx_0'} \over {(2E + U(x_0')^2)^{\half}}} = t \ ,
\eqno(2.38)
$$
where $x_0 = X_0$ at $t = 0$.

Given $x_0(t)$, and hence $U'(x_0(t))$, we can solve the linear
equations for $a_1, a_2, b_1, b_2$. Of course, one solution is that
these four functions all vanish. The supersymmetry variations (2.21)
and (2.22) suggest the solution
$$
\eqalignno{
a_1 &= \l \xdot_0 + \m U &(2.39a) \cr
a_2 &= \l U + \m \xdot_0 &(2.39b) \cr
b_1 &= \s \xdot_0 + \t U &(2.39c) \cr
b_2 &= \s U + \t \xdot_0 \ , &(2.39d) \cr
}
$$
where $\l, \m, \s, \t$ are arbitrary constants, and $U$ denotes
$U(x_0(t))$. These functions do satisfy the
equations of motion, e.g.
$$
\eqalign{
\adot_1 &= \l {\ddot x}_0 + \m U' \xdot_0 \cr
&= \l UU' + \m U' \xdot_0 \cr 
&= U' a_2 \ , }
\eqno(2.40)
$$ 
and the presence of four constants implies that (2.39) is the general
solution. The value of the conserved supersymmetry charge $\Qa$ is
$$
\eqalign{
\Qa &= \xdot_0 b_1 - Ub_2 \cr
&= \s (\xdot_0^2 - U^2) \cr
&= 2E\s \ , }
\eqno(2.41)
$$
and similarly $\Qb = -2E\l$, $\tQa = 2E\t$ and $\tQb = -2E\m$. The $R$
charges take the values $R_a = E(\s^2 - \t^2)$, $R_b = E(\l^2 - \m^2)$
and $R = 2E(\l\t - \m\s)$. There is a problem, however, if $E = 0$, for then 
$$
\xdot_0 = \pm U
\eqno(2.42)
$$
and the expressions (2.39) depend on only two arbitrary
constants. Eq.(2.42) is the Bogomolny equation for this system. We postpone
discussion of the general solution in this case to the next Section.

The remaining equation for $x_1$ is the inhomogeneous linear equation
$$
{\ddot x}_1 = (UU')' x_1 + 2E(\l\t - \m\s)U'' \ , 
\eqno(2.43)
$$
where we have substituted the conserved value of $R = a_1 b_2 - a_2 b_1$.
The supersymmetry transformations and the mini-time-translation
suggest that solutions can be constructed from $U$ and $\xdot_0$. It
may be verified, using (2.18a) and (2.36), that a particular integral 
of (2.43) is
$$
x_1 = (\l\t - \m\s)U \ .
\eqno(2.44)
$$
A solution of the homogeneous equation ${\ddot x}_1 = (UU')' x_1$ is
$x_1 = C_1 \xdot_0$, with $C_1$ a constant, since ${d^3x_0 \over
dt^3} = {d(UU') \over dt} = (UU')' \xdot_0$. A second solution must satisfy
$\xdot_1 \xdot_0 - x_1 {\ddot x}_0 = C_2$ for some constant
(Wronskian) $C_2$. Write $x_1 = f(t)\xdot_0$. Then $f$ must satisfy
${\dot f} = C_2 / \xdot_0^2$, so
$$
{df \over dx_0} = {C_2 \over \xdot_0^3}
= {C_2 \over {(2E + U^2)^{3 \over 2}}} \ ,  
\eqno(2.45)
$$
and hence the second solution is
$$
x_1 = C_2 (2E + U^2)^{\half} \int_{X_0}^{x_0(t)} 
{{dx_0'} \over {(2E + U(x_0')^2)^{3 \over 2}}} \ .
\eqno(2.46)
$$ 
The complete solution of (2.43) is therefore
$$
x_1 = (\l\t - \m\s)U + C_1 (2E + U^2)^{\half} + 
C_2 (2E + U^2)^{\half} \int_{X_0}^{x_0(t)} 
{{dx_0'} \over {(2E + U(x_0')^2)^{3 \over 2}}} \ .
\eqno(2.47)
$$
The value of the energy constant $H_1$ is $C_2$.

We have therefore found the general solution of the equations of
motion (2.18), in terms of eight constants of integration $X_0, E, \l,
\m, \s, \t, C_1, C_2$. Our solution is incomplete, however, if $E =
0$.

We conclude this Section with a brief discussion of the supersymmetry
algebra and how it is realized on the dynamical variables. In the
model considered here there are two supersymmetry operators $Q$ and
$\tQ$ (we use the same notation as for the associated conserved
charges). Together with ${d \over dt}$ they are a basis for a super
Lie algebra over the reals with nontrivial relations
$$
Q^2 = {d \over dt} \ , \ \tQ^2 = -{d \over dt} \ , \ Q\tQ + \tQ Q = 0 \ .
\eqno(2.48)
$$
Formally, the algebra has a representation on the dynamical variables
$$
Q x = \p_1 \ , \ Q \p_1 = \xdot \ , \ Q \p_2 = U
\eqno(2.49a)
$$
$$
\tQ x = \p_2 \ , \ \tQ \p_2 = -\xdot \ , \ \tQ \p_1 = -U \ .
\eqno(2.49b)
$$
This ``on shell' representation requires that the equations 
$\pdot_1 = U'\p_2$ and
$\pdot_2 = U'\p_1$ are satisfied, so that, for example $Q^2\p_2 = QU =
U'(Qx) = U'\p_1 = \pdot_2$. $Q$, $\tQ$ and ${d \over dt}$ are
all symmetries of the Lagrangian, provided $Q$ and $\tQ$ are treated
as antiderivations (an extra minus sign in the Leibniz rule when $Q$
or $\tQ$ goes past a fermionic variable, e.g. $Q(\p_1\p_2) =
(Q\p_1)\p_2 - \p_1 Q\p_2 = \xdot \p_2 -\p_1 U$). Now although the 
actions of $Q$ and $\tQ$ given by (2.49) make formal
sense, they cannot be regarded as variations of the dynamical
variables. A bosonic variable cannot be varied by a fermionic
one. Moreover, the vague requirement that the coefficients of $Q$ and
$\tQ$ should be anticommuting, common in the literature, is not
sufficiently precise. However, requiring the coefficients to be
elements of $B_o$ is sufficiently precise, and leads to eqs.(2.5) and
(2.9) as genuine variations.

The super Lie algebra over the reals becomes an ordinary Lie algebra
if the coefficients lie in $B$, with $Q$ and $\tQ$ having
coefficients in $B_o$ and ${d \over dt}$ having a
coefficient in $B_e$. With $B$ generated by $\a$ and $\b$, this real
Lie algebra is six dimensional, with generators 
$$
\Qa \ , \ \Qb \ , \ \tQa  \ , \ \tQb \ , \ {d \over dt} \ , \
{\widetilde {d \over dt}} \ ,
\eqno(2.50)
$$
where ${d \over dt}$ is the usual time derivative and ${\widetilde {d
\over dt}}$ the mini-time-derivative. Almost all these generators
commute, except that 
$$
[\Qa \ , \ \Qb] = -2{\widetilde {d \over dt}} \ , \ \ \ 
[\tQa \ , \ \tQb] = 2{\widetilde {d \over dt}} \ .
\eqno(2.51)
$$
${d \over dt}$ is a central element which acts in the obvious way.
The action of the other generators is given by
(2.21), (2.22) and (2.28). So rather than think of the super Lie algebra as
an extension of the one-dimensional Lie algebra with generator 
${d \over dt}$, one may regard it as shorthand
for a larger Lie algebra with a particular structure related to
$B$. There is an infinite family
of ordinary Lie algebras, one for each choice of $B$, all of which
stem from the same super Lie algebra. This interpretation of a
super Lie algebra as a family of ordinary Lie algebras is discussed by
Freund [2].

The Lie group generated by the six elements (2.50)
is the true symmetry group of our system, the supergroup. 
From the infinitesimal action on the constants of 
integration of the general solution, it is clear that the supergroup 
has six-dimensional orbits in the space of
solutions. Only $E$ and $C_2$ are invariant. 

{\bf III. Zero Energy Solutions}

When the energy $E = 0$, the method described above does not
give the general solution of the equations of motion (2.18). For this
value of $E$
$$
\xdot_0^2 - U^2 = 0 \ ,
\eqno(3.1)
$$
so $x_0$ satisfies the first order Bogomolny equation
$$
\xdot_0 = \pm U \ .
\eqno(3.2)
$$
For either choice of sign, $\xdot_0$ and $U$ are no longer independent
functions of time, so the expressions (2.39) depend effectively on only
two arbitrary constants, and are no longer the general solution.

For simplicity, let us choose the upper sign in (3.2). The lower sign
choice is essentially the same, and corresponds to a time reversal.
Then the solution of (3.2) is
$$
\int_{X_0}^{x_0} {{dx_0'} \over {U(x_0')}} = t \ .
\eqno(3.3)
$$

To find the general solution of the equations for $a_1, a_2, b_1,
b_2$, it helps to consider the limit $E \rightarrow 0$ of the solution
given earlier. Note that for small non-zero $E$,
$$
\eqalign{
\xdot_0 &= (2E + U^2)^{\half} \cr
&=  U + {E \over U} + O(E^2) \ . } 
\eqno(3.4)
$$
A suitable linear combination of $\xdot_0$ and $U$ is
proportional to ${1 \over U}$ in the limit $E \rightarrow 0$. We
therefore try
$$
a_1 = {\l \over U} + \m U \ .
\eqno(3.5)
$$
Then
$$
\adot_1 = -{\l \over U^2}U' \xdot_0 + \m U' \xdot_0 = U' (-{\l \over
U} + \m U)
\eqno(3.6)
$$
if $\xdot_0 = U$. Thus $a_2 = -{\l \over U} + \m U$ gives a solution
of (2.18c), and it is easily checked that (2.18d) is also
satisfied. Similarly we can solve eqs.(2.18e) and (2.18f). So the
general solution of eqs.(2.18c-f) is
$$
\eqalign{
a_1 &= {\l \over U} + \m U \ , \ a_2 = -{\l \over U} + \m U \cr
b_1 &= {\s \over U} + \t U \ , \ b_2 = -{\s \over U} + \t U  \ , } 
\eqno(3.7)
$$
where $\l, \m, \s, \t$ are arbitrary constants.

The constants of the motion take the
following values 
$$
\eqalign{
\Qa &= -\tQa = 2\s \ , \ -\Qb = \tQb = 2\l \ , \cr
R_a &= 2\s\t \ , \ R_b = 2\l\m  \ , \ R = 2(\l\t - \m\s) \ .} 
\eqno(3.8)
$$
These values are generally nonzero because of the careful way the
limit $E \rightarrow 0$ was taken, even though previously
these quantities were proportional to $E$.

The remaining equation for $x_1$ also needs special treatment. This 
equation is
$$
{\ddot x}_1 = (UU')' x_1 + RU''
\eqno(3.9)
$$
where $R$ is the constant given in (3.8). The previous solution had a
particular integral proportional to $U$, and one homogeneous solution 
proportional to $\xdot_0$. When $E = 0$, and $\xdot_0 = U$, one
homogeneous solution is still $U$. But a new particular integral is
required. Again the limiting procedure suggests that this should be 
proportional to ${1 \over U}$, and this is correct. Finding a second
homogeneous solution is as before, but with $E = 0$. The result
is that the general solution of (3.9) is
$$
x_1 = -{R \over 2U} + C_1 U + C_2 U \int_{X_0}^{x_0(t)} 
{{dx_0'} \over U(x_0')^3} \ ,
\eqno(3.10)
$$
where $C_1$ and $C_2$ are arbitrary constants. $H_1 = C_2$, as before.

Note that in the zero energy, Bogomolny case, the orbits of the supergroup
on the space of solutions are four-dimensional, rather than six-dimensional. 
Only the coefficients of $U$ in
(3.7) and (3.10) can be varied by the group action. This is consistent with
the observation that the supersymmetry generator $\d + \td$ produces
no variation at all when $\xdot_0 = U$ and $a_1, a_2, b_1, b_2, x_1$
all vanish.

{\bf IV. $N=1$ Supersymmetric Mechanics}

Another example of a solvable supersymmetric mechanical model is that
of a particle moving in one dimension with $N=1$ supersymmetry
(sometimes referred to as $N = \half$ supersymmetry) [4,7]. The
supersymmetry algebra is simply $Q^2 = {d \over dt}$. The dynamical
variables are a bosonic variable $x(t)$ and a single fermionic
variable $\p (t)$, taking values in $B_e$ and $B_o$ respectively. The
Lagrangian is
$$
L = \half \xdot^2 + \half \pdot\p + \a U(x) \p \ .
\eqno(4.1)
$$
$\a$ is an odd constant, an
element of $B_o$. It is necessary for $\a$ to be odd, and $L$
even, otherwise the equations of motion are contradictory.
This model is a variant of the usual nontrivial $N=1$
supersymmetric mechanical models. Normally, such a model has two or
more fermionic variables [8]. Here, one of these is replaced by the odd
constant $\a$. 

Taking the variation of $L$, ignoring total time derivatives, and
shifting the variations to the left, gives
$$
\D L = -\D x ({\ddot x} - \a U' \p) - \D \p (\pdot + \a U) \ ,
\eqno(4.2)
$$
so the equations of motion are
$$
\eqalignno{
{\ddot x} &= \a U' \p &(4.3a) \cr
\pdot &= -\a U \ . &(4.3b) \cr
}
$$
We see that both sides of eq.(4.3a) are in $B_e$, and both sides of
(4.3b) in $B_o$.

The supersymmetry variations are 
$$
\d x = \e \p \ , \ \d \p = \e \xdot \ ,
\eqno(4.4)
$$
where $\e$ is an arbitrary infinitesimal odd constant. The
corresponding variation of $L$ is
$$
\d L = \e(\half \xdot\pdot + \half {\ddot x} \p - \a U \xdot) \ .
\eqno(4.5)
$$
Let us introduce $V(x)$, satisfying $V' = U$. Then we can write $\d L$
as a total time derivative
$$
\d L = \e {d \over dt}(\half \xdot \p - \a V) \ .
\eqno(4.6)
$$
Hence $L$ is supersymmetric, and the conserved supersymmetry charge is
$$
Q = \xdot \p + \a V
\eqno(4.7)
$$
Using standard arguments, we also obtain the energy
$$
H = \half \xdot^2 - \a U \p \ .
\eqno(4.8)
$$
Its conservation follows from the equations of motion, together with
$\a^2 = 0$.

We may again obtain a concrete realization of this model by supposing 
that the Grassmann algebra $B$ has
just two generators. Without loss of generality we may suppose that $\a$
is one of these generators, and that the other is $\b$. The algebra is
then identical to that in Section II. Note that if $B$ had only one
generator, then $\a\p$ would be zero, and the model would become
trivial.

We write the component expansion of the dynamical variables as
$$ 
\eqalignno{
x(t) &= x_0(t) + x_1(t)\a\b &(4.9a) \cr
\p(t) &= a(t)\a + b(t)\b &(4.9b) \cr
}
$$
where $x_0,x_1,a,b$ are ordinary functions.
The Lagrangian has the expansion $L = L_0 + L_1\a\b$,
where
$$
\eqalignno{
L_0 &= \half {\xdot}_0^2  &(4.10a) \cr
L_1 &= \xdot_0 \xdot_1 + \half \adot b - \half \bdot a + U(x_0)b
\ . &(4.10b) \cr
}
$$
The equations of motion become
$$ 
\eqalignno{
{\ddot x}_0 &= 0 &(4.11a) \cr
{\ddot x}_1 &= U'(x_0) b &(4.11b) \cr
\adot &= -U(x_0) &(4.11c) \cr
\bdot &= 0 \ . &(4.11d) \cr
}
$$
These can be obtained as the components of eqs.(4.3). They are also
the variational equations obtained from $L_1$ and $L_0$, and, as
before, $L_0$ is redundant.

The equations (4.11) imply the conservation of
$$ 
\eqalignno{
\Qa &= {\xdot}_0 a + V(x_0) &(4.12a) \cr
\Qb &= {\xdot}_0 b &(4.12b) \cr
H_0 &= \half {\xdot}_0^2 &(4.12c) \cr
H_1 &= \xdot_0 \xdot_1 - U(x_0) b \ , &(4.12d) \cr
}
$$
and these are the components of $Q$ and $H$.

It is straightforward to solve the equations (4.11), starting with
$$ 
x_0 = \l t + \m \ , \ b = \n 
\eqno(4.13)
$$
where $\l, \m, \n$ are arbitrary constants. The energy $H_0$ is $\half
\l^2$. We now regard
$\Qa$ as a constant of integration, obtaining
$$
a = {\Qa \over \l} - {1 \over \l}V(\l t + \m)
\eqno(4.14)
$$
as the solution of (4.11c). Finally, treating
$H_1$ similarly, we have
$$
\xdot_1 = {H_1 \over \l} + {\n \over \l}U(\l t + \m)
\eqno(4.15)
$$
so
$$
x_1 = {H_1 \over \l}t + X_1 + {\n \over \l^2}V(\l t + \m)
\eqno(4.16)
$$
where $X_1$ is a constant. The general solution of the model involves
six arbitrary constants $\l, \m, \n, \Qa, H_1, X_1$. 

We obtain a genuine Lie algebra of symmetries from the components 
of the supersymmetry transformation and time translation. Starting 
with $\d$ we obtain two independent supersymmetry variations
$$ 
\eqalignno{
\da x &= \e \a \p \ , \ \da \p = \e \a \xdot &(4.17a) \cr
\db x &= \e \b \p \ , \ \db \p = \e \b \xdot \ , &(4.17b) \cr
}
$$
where $\e$ is now infinitesimal and real. Writing $x$ and $\p$ in
terms of components, we find
$$ 
\eqalignno{
\da x_1 &= \e b \ , \ \da a = \e \xdot_0 &(4.18a) \cr
\db x_1 &= -\e a \ , \ \db b = \e \xdot_0 \ , &(4.18b) \cr
}
$$
with all other variations vanishing. These are symmetries of $L_1$,
and trivially of $L_0$. In addition there is symmetry under an
infinitesimal time translation
of all the dynamical quantities. Finally, there is symmetry under
the infinitesimal mini-time-translation
$$
\D x_1 = \e \xdot_0 \ .
\eqno(4.19)
$$

The supergroup of this system therefore has generators
$$
\Qa \ , \ \Qb \ , \ {d \over dt} \ , \
{\widetilde {d \over dt}} \ ,
\eqno(4.20)
$$
where the action of $\Qa , \Qb$ is defined by (4.18), 
${d \over dt}$ is the time derivative, and 
${\widetilde {d \over dt}}$ the mini-time-derivative defined by (4.19). 
The only non-trivial bracket is
$$
[\Qa \ , \ \Qb] = -2{\widetilde {d \over dt}} \ . 
\eqno(4.21)
$$
Acting with the supergroup we may vary $\m, \nu, \Qa, X_1$,
but not the constants defining the energy $\l$ and $H_1$.
 
The solution as we have presented it doesn't make sense if $\l =
0$. This is the zero energy, Bogomolny case. 
If $H_0 = 0$ then $\xdot_0 = 0$, so $x_0$ takes a constant
value $\m$, hence $U$ and $U'$ take constant values $U(\m)$ and
$U'(\m)$. The general solution is then easily found to be
$$
\eqalignno{
x_0 &= \m \ , \ b = \n &(4.22a) \cr
a &= -U(\m) (t - t_0) &(4.22b) \cr
x_1 &= \half U'(\m) \n t^2 +rt + X_1 &(4.22c) \cr
}
$$
where $\m, \n, t_0, r, X_1$ are constants of integration. The second
energy constant is $H_1 = -U(\m)\n$. Supersymmetry transformations
and time translations change the constants $r$, $X_1$ and $t_0$. 
However, unlike in the $H_0 \neq 0$ case, eq.(4.18b)
implies that the value of $b$ cannot be changed, and the orbits of the
supergroup are three-dimensional rather than four-dimensional. 

{\bf V. Conclusions}

We have presented two supersymmetric classical mechanical models. By supposing
that the dynamical variables take values in a Grassmann algebra $B$ with two
generators, we have deconstructed the models into component form and 
obtained equations of motion which can be
explicitly solved. These equations are the variational equations of a
Lagrangian $L_1$ of non-standard form, and in each case, the ``body'' 
variable $x_0$ obeys a classical equation unaffected by the fermionic 
variables. A genuine Lie group, generated from the supersymmetry 
algebra, acts on the space of solutions. 

One could ask how the solutions would look if the dynamical variables were
reconstructed to be $B$-valued, or further combined into superspace
dynamical variables. At first sight there is only a slight gain in
elegance, but this needs more careful study. It is also of interest to
know whether the equations remain solvable if $B$ is a larger algebra. 

The model discussed in Section IV involved an odd constant
$\a$. Possibly, Grassmann-valued constants are of use in other
supersymmetric models. For example, it might be possible in certain
``brane'' models to have a non-real cosmological constant.

One of the motivations for this work was to better understand the
solitons that occur in many supersymmetric field theories. These are
solutions of the classical field equations, with the fermionic fields
set equal to zero. They usually also satisfy first-order Bogomolny
equations. It would be much more satisfactory if they
could be regarded as special cases of solutions where the
fermionic fields are nonzero. Our mechanical models suggest that the
''body'' fields of the soliton will be unaffected by the 
fermionic fields. But the general solutions will involve nonzero fermionic
fields coupled to the soliton, and in addition there will be nonzero
bosonic fields with values in the even, non-real part of the Grassmann
algebra. 

The connection between the classical models discussed here and their 
quantized versions is also worth exploring. The 
Heisenberg equations of the quantized theory may
be formally the same as the equations that we have solved, but
$x,\xdot$ and $\p,\pdot$ need to obey canonical commutation and
anticommutation relations, respectively. It would be interesting to
know whether the general classical solution describes a suitable limit
of a quantum state.

\vskip 40pt
{\bf Acknowledgements}

I am grateful to Alan Macfarlane and Jonathan Evans for discussions 
during the course of this work.

\vfill\eject
{\bf References}

\item {1.} B. De Witt, {\it Supermanifolds, 2nd ed.} 
(Cambridge University Press, Cambridge, 1992).

\item {2.} P.G.O. Freund, {\it Introduction to Supersymmetry} 
(Cambridge University Press, Cambridge, 1986). 

\item {3.} F.A. Berezin and M.S. Marinov, Ann. Phys. {\bf 104}, 336 (1977).

\item {4.} R. Casalbuoni, Nuovo Cim. {\bf 33A}, 389 (1976).

\item {5.} E.B. Bogomolny, Sov. J. Nucl. Phys. {\bf 24}, 449 (1976).

\item {6.} E. Witten, Nucl. Phys. {\bf B188}, 513 (1981).

\item {7.} M. Sakimoto, Phys. Lett. {\bf B151}, 115 (1985).

\item {8.} A.J. Macfarlane, Nucl. Phys. {\bf B438}, 455 (1995).
\bye